\documentclass{article}

\usepackage{amssymb,amsfonts,amsthm}
\usepackage[cp1251]{inputenc}
\title{Generalized Hermitian Codes over $GF(2^r)$}
\author{S.V.Bulygin}

\newtheorem{prop11}{Proposition}[section]
\newtheorem{th12}[prop11]{Theorem}
\newtheorem{rem13}[prop11]{Remark}
\newtheorem{prop14}[prop11]{Proposition}

\newtheorem{rem21}{Remark}[section]
\newtheorem{rem23}[rem21]{Remark}
\newtheorem{lem22}[rem21]{Lemma}
\newtheorem{th24}[rem21]{Theorem}
\newtheorem{def25}[rem21]{Definition}
\newtheorem{def26}[rem21]{Definition}
\newtheorem{prop27}[rem21]{Proposition}
\newtheorem{th28}[rem21]{Theorem}
\newtheorem{rem29}[rem21]{Remark}

\newtheorem{def31}{Definition}[section]
\newtheorem{def32}[def31]{Definition}
\newtheorem{prop33}[def31]{Proposition}
\newtheorem{prop34}[def31]{Proposition}
\newtheorem{th35}[def31]{Theorem}
\newtheorem{th36}[def31]{Theorem}
\newtheorem{prop37}[def31]{Proposition}
\newtheorem{prop38}[def31]{Proposition}
\newtheorem{ex39}[def31]{Example}

\newtheorem{prop41}{Proposition}[section]
\newtheorem{th42}[prop41]{Theorem}
\newtheorem{lem43}[prop41]{Lemma}
\newtheorem{lem44}[prop41]{Lemma}
\newtheorem{cor45}[prop41]{Corollay}

\begin{document}
\maketitle
    \section{Introduction and preparation results}
    The main focus of the present research is on construction of codes
    on function fields, which we called Generalized Hermitian function fields
    (GH-fields). The term follows from the fact that well-known
    Hermitian function fields are a special case of the considered family. We
    first will introduce some preparation material (notions,
    theorems etc.). The next section 2 is a technical core of
    this work, which will give us an opportunity to calculate or
    at least estimate parameters of codes, constructed on
    GH-fields (Generalized Hermitian codes; GH-codes) in section
    3. These codes turn out to have nice properties similar to
    those of Hermitian codes, but over larger alphabet. In fact
    some of these codes over $\mathbb{F}_{8}$ attain record values
    for given parameters; one code delivers a new record. Also their generator matrices can
    be effectively constructed. Section 4 is devoted to investigating a duality property of GH-codes. It turns out that
    the duality property of GH-codes is analogous to that of Hermitian codes. Section 5 gives some specific computational
    results. We finish with conclusions and acknowledgements in sections 6 and 7.
    \\
    \indent
    So, first of all recall that Hermitian function fields are from
    the family of elementary abelian $p$-extensions of $K(x)$, where
    $char K=p>0$. The main properties of these function fields
    that are of importance for coding theory are collected in the
    following (Lemma VI.4.4,\cite{1}):
    \begin{prop11}
        The Hermitian function field over $\mathbb{F}_{q^2}$, $q$ is a prime power, can
        be defined by
        \begin{equation}
            H=\mathbb{F}_{q^2}(x,y) \: with \: y^q+y=x^{q+1}.
            \label{1-1}
        \end{equation}
        It has the following properties:
        \\
        (a) The genus of $H$ is $g=q(q-1)/2$.\\
        (b) $H$ has $q^3+1$ places of degree one over
        $\mathbb{F}_{q^2}$, namely\\
        \indent (1) the common pole $Q_{\infty}$ of $x$ and $y$,
        and\\
        \indent (2) for each $\alpha\in\mathbb{F}_{q^2}$, there
        are $q$ elements $\beta\in\mathbb{F}_{q^2}$ such that
        $\beta^q+\beta = \alpha^{q+1}$, and for all such pair
        $(\alpha,\beta)$ there is a unique place
        $P_{\alpha,\beta}\in\mathbb{P}_{H}$ of degree one with
        $x(P_{\alpha,\beta})=\alpha$ and
        $y(P_{\alpha,\beta})=\beta$.\\
        (c) $H/ \mathbb{F}_{q^2}$ is a maximal function field.\\
        (d) For $r\ge 0$, the elements $x^i y^j$ with $0\le i,0\le
        j\le q-1$ and $iq+j(q+1)\le r$ form a basis of
        $\mathcal{L}(rQ_{\infty})$.
    \end{prop11}\label{prop11}
    Now we present a family of function fields, which we call
    \emph{GH-fields}. The following theorem is from \cite{2}.
    \begin{th12}
        Let $r\ge 2$. Then the curve
        \begin{equation}
            y^{q^{r-1}}+\dots+y^q+y = x^{1+q}+x^{1+q^2}+\dots+x^{q^{r-2}+q^{r-1}}
            \label{1-2}
        \end{equation}
        over $\mathbb{F}_{q^r}$ is absolutely irreducible. The
        corresponding function field $F/\mathbb{F}_{q^r}$ of this curve has genus
        \begin{displaymath}
            g = q^{r-1}(q^{r-1}-1)/2,
        \end{displaymath}
        and the number of rational places is
        \begin{displaymath}
            N = 1+q^{2r-1}.
        \end{displaymath}
    \end{th12}\label{th12}
    \begin{rem13}
        (i) Note that we can write equation (\ref{1-2}) as
        $s_{r,1}(y) = s_{r,2}(x)$, where $s_{r,1}(y)$ and
        $s_{r,2}(x)$ are the first and the second symmetric
        polynomials of $(y,y^q,\dots,y^{q^{r-1}})$ and
        $(x,x^q,\dots,x^{q^{r-1}})$ respectively.\\
        (ii) For $r=2$ this curve is the Hermitian curve over
        $\mathbb{F}_{q^2}$, and the result is well known
        (Proposition 1.1).
    \end{rem13}
    Our aim is to present an analogue of Proposition 1.1 for
    GH-fields from Theorem 1.2.
    \\
    \indent
    Basically, (a) from Proposition 1.1 has its analogue in
    Theorem 1.2, (c) does not hold for GH-curves starting with
    $r\ge 3$. So, we have to find the analogue for (b) and (d).
    The answer to (d) is very important for construction of
    GH-codes and is given in Theorem 2.8 from section 2, but in
    order to justify the proof of this theorem we need some
    preparation.
    \begin{prop14}
        Let $F/\mathbb{F}_{q^r}$ be a function field of the curve
        defined by (\ref{1-2}). Then the following holds:
        \\
        \indent
        (a) The pole $P_{\infty}\in\mathbb{P}_{\mathbb{F}_{q^r}(x)}$ (by $\mathbb{P}_F$ we denote the set of place of
    a function field $F$)
        of $x$ in $\mathbb{F}_{q^r}(x)$ has a unique extension
        $Q_{\infty}\in\mathbb{P}_F$, and $Q_{\infty}|P_{\infty}$
        is totally ramified (i.e.
        $e(Q_{\infty}|P_{\infty})=q^{r-1}$). Hence $Q_{\infty}$ is
        a place of $F/\mathbb{F}_{q^r}$ of degree one.
        \\
        \indent
        (b) The pole divisor of $x$ is
        $(x)_{\infty}=q^{r-1}Q_{\infty}$, and of $y$ is
        $(y)_{\infty}=(q^{r-1}+q^{r-2})Q_{\infty}$.
        \\
        \indent
        (c) For each $\alpha\in\mathbb{F}_{q^r}$, there are
        $q^{r-1}$ elements $\beta\in\mathbb{F}_{q^r}$ such that
        $\beta^{q^{r-1}}+\dots+\beta=\alpha^{q^{r-1}+q^{r-1}}+\dots+\alpha^{1+q}=:f(\alpha)$,
        and for all such pairs $(\alpha,\beta)$ there is a unique
        place $P_{\alpha,\beta}\in\mathbb{P}_{F}$ of degree one
        with $x(P_{\alpha,\beta})=\alpha$ and
        $y(P_{\alpha,\beta})=\beta$.
    \end{prop14}\label{prop14}
    \begin{proof}
        (a) It follows from the proof of Theorem 4.1 , \cite{2} (it
        is Theorem 1.2 in our text).
        \\
        \indent
        (b) $(x)_{\infty}=q^{r-1}Q_{\infty}$ follows from (a)
        (cf.Theorem I.4.11, \cite{1}). By (\ref{1-2}) $x$ and $y$
        have the same poles, hence $Q_{\infty}$ is the only pole
        of $y$ as well. As
        $q^{r-1}v_{Q_{\infty}}(y)=v_{Q_{\infty}}(y^{q^{r-1}}+\dots+y)=v_{Q_{\infty}}
        (x^{q^{r-1}+q^{r-2}}+\dots+x^{1+q})=q^{r-1}(q^{r-1}+q^{r-2})$
        we obtain $(y)_{\infty}=(q^{r-1}+q^{r-2})Q_{\infty}$.
        \\
        \indent
        (c) For the first part of the statement see the proof of
        Theorem 4.1. from \cite{2}. Now, suppose there is some
        $\beta\in\mathbb{F}_{q^r}$ such that
        $\beta^{q^{r-1}}+\dots+\beta=f(\alpha)$. It follows that
        $(\beta+\gamma)^{q^{r-1}}+\dots+(\beta+\gamma)=f(\alpha)$
        for all $\gamma$ with $\gamma^{q^{r-1}}+\dots+\gamma=0$,
        so
        \begin{displaymath}
            T^{q^{r-1}}+\dots+T-f(\alpha)=\prod_{j=1}^{q^{r-1}}(T-\beta_j)
        \end{displaymath}
        with pairwise distinct elements
        $\beta_i\in\mathbb{F}_{q^r}$. By Corollary III.3.8(c) from
        \cite{1}, there exists for $j=1,\dots,q^{r-1}$ a unique
        place $P_j\in\mathbb{P}_F$ such that $P_j|P_{\alpha}$ and
        $y-\beta_j\in P_j$, and the degree of $P_j$ is one, which
        means $x(P_j)=\alpha, y(P_j)=\beta$.
    \end{proof}
    In the proof we have used some methods from the proof of
    Proposition VI.4.1 (\cite{1}). Now using this proposition we
    go on to the question of the structure of vector spaces
    $\mathcal{L}(sQ_{\infty})$ for given $s$.

    \section{Structure of a Weierstrass semigroup of the place at infinity}
    Our aim now is to determine the basis of
    $\mathcal{L}(sQ_{\infty})$ for given $s$. This problem is
    closely connected with finding a Weierstrass semigroup of
    $Q_{\infty}$ up to a given parameter $s$.
    \\
    \indent
    In the case of Hermitian curves the situation is quite simple
    as Weierstrass semigroup is generated by the orders at infinity of $x$ and
    $y$. This gives rise to the fact that
    $\mathcal{L}(sP_{\infty})$ is generated by functions of the
    form $x^i y^j$, where $q i+(q+1)j\le s,$ where $q, (q+1)$ are orders
    of $x$ and $y$ respectively.
    Our situation is more complicated. We will restrict ourselves
    to the case $q=2$. So we are considering a curve
    \begin{equation}
        y^{2^{r-1}}+\dots+y^2+y = x^{2^{r-1}+2^{r-2}}+\dots+x^3
        \label{2-1}
    \end{equation}
    over $\mathbb{F}_{2^r}, r\ge 3$.
    \begin{rem21}
        Our case does not include all curves from the considered
        family over the field of characteristics 2. Indeed, we can
        consider $\mathbb{F}_{2^{r k}}$ as $\mathbb{F}_{(2^r)^k}$,
        so the equation will be
        \begin{displaymath}
            y^{(2^r)^{k-1}}+\dots+y^{2^r}+y =
            x^{(2^r)^{k-1}+(2^r)^{k-2}}+\dots+x^{1+2^r}
        \end{displaymath}
        as oppose to
        \begin{displaymath}
            y^{2^{r k-1}}+\dots+y^2+y =
            x^{2^{r k-1}+2^{r k-2}}+\dots+x^3
        \end{displaymath}
        in the case $q=2$ (constant field is $\mathbb{F}_{2^{r
        k}}$).
        \\
        \indent
        If we consider all $q$ and $r$ such that $q^r$ is fixed
        and $q$ is a power of 2, then the maximal number of
        rational points will be when $q=2$. This follows from the
        fact that $N=1+q^{2r-1}=1+\frac{(q^r)^2}{q}$. So as $q$
        grows $N$ decreases. The same argument, of course, works for
        arbitrary characteristics.
        \\
        \indent
        However, the ratio
        \begin{displaymath}
            \frac{N}{g}=\frac{2(1+q^{2r-1})}{q^{r-1}(q^{r-1}-1)}=
            \frac{2(1+\frac{(q^r)^2}{q})}{\frac{q^r}{q}(\frac{q^r}{q}-1)}
        \end{displaymath}
        is the lowest when $q=2$, and grows as $q$ goes up. That
        is why studying the curves over $\mathbb{F}_{q^r}$, where
        $q$ is a power of a prime is of interest.
    \end{rem21}
    As a prelude to finding the Weierstrass semigroup of
    $Q_{\infty}$ and corresponding basis of
    $\mathcal{L}(sQ_{\infty})$, let us first find orders of some
    functions at infinity (i.e. at $Q_{\infty}$). The following
    lemma will give us an opportunity to find numbers that
    generate the whole Weierstrass semigroup of
    $Q_{\infty}$.
    \begin{lem22}
        If we denote $ord(f):=-v_{Q_{\infty}}(f), \epsilon:=x^3+y^2, \theta:=
        \epsilon+x y$, then the following hold:
          \begin{itemize}
            \item $ord(x)=2^{r-1};$
            \item $ord(y)=2^{r-1}+2^{r-2};$
            \item $ord(\epsilon)=ord(x y);$
            \item $ord(\theta)=2^r+1.$
          \end{itemize}
    \end{lem22}\label{lem22}

    \begin{proof}
    As was noted before, we have that $x$ and $y$ have orders at
    infinity respectively $ord(x)=-v_{Q_{\infty}}(x)=2^{r-1}$ and
    $ord(y)=-v_{Q_{\infty}}(y)=2^{r-1}+2^{r-2}$. When $r\ge 3$ we
    have that $ord(x)$ and $ord(y)$ are both even, so we cannot
    hope on them to generate the Weierstrass semigroup. So we have
    to search for other generator(s).
    \\
    \indent
    As we see $ord(x^3)=ord(y^2)=2^r+2^{r-1}=3\cdot 2^{r-1}$. If
    we consider their sum $x^3+y^2$ we may hope that
    $ord(x^3+y^2)$ will be lower than $ord(x^3)=ord(y^2)$ and
    differ from those orders that could be generated by $ord(x)$
    and $ord(y)$.
    So let us put $\epsilon:=x^3+y^2$. By squaring both sides of
    (\ref{2-1}) we have
    \begin{equation}
        y^{2^r}+\dots+y^4+y^2 = x^{2^r+2^{r-1}}+\dots+x^6.
        \label{2-2}
    \end{equation}
    Now, considering that $\alpha=-\alpha$ in a field of
    characteristics 2 we have
    \begin{eqnarray}
        \epsilon^{2^{r-1}} = x^{3\cdot 2^{r-1}}+y^{2^r} =
        x^{2^r+2^{r-1}}+y^{2^r} = |using \ (2)|
        \nonumber\\
        =y^{2^{r-1}}+\dots+y^4+y^2+x^{2^r+2^{r-2}}+\dots+x^6.
        \nonumber
    \end{eqnarray}
    We then use the fact that
    \begin{displaymath}
        y^{2^{r-1}}+\dots+y^4+y^2 =
        y+x^{2^{r-1}+2^{r-2}}+\dots+x^3.
    \end{displaymath}
    So,
    \begin{displaymath}
        \epsilon^{2^{r-1}} =
        y+(x^{2^{r-1}+2^{r-2}}+\dots+x^3)+(x^{2^r+2^{r-2}}+\dots+x^6).
    \end{displaymath}
    All summands of the form $x^{2^i+2^j}, i,j>0$ from the first
    bracket will be canceled, as
    $x^{2^i+2^j}=(x^{2^{i-1}+2^{j-1}})^2$. So the first bracket
    reduces to $x^{2^{r-1}+1}+x^{2^{r-2}+1}+\dots+x^3$.
    Analogously, in the second bracket all summands of the form $x^{2^i+2^j}; 0\le j<i\le
    r-1$ will be crossed out (as they are present in the first
    bracket), so the second bracket reduces to
    $x^{2^r+2^{r-2}}+x^{2^r+2^{r-3}}+\dots+x^{2^r+1}$. After
    reducing we have
    \begin{displaymath}
        \epsilon^{2^{r-1}} =
        y+x^{2^r+2^{r-2}}+\dots+x^{2^r+2}+x^{2^{r-1}+1}+x^{2^{r-2}+1}+\dots+x^3.
    \end{displaymath}
    Orders (at $Q_\infty$) of all summands here are pairwise
    distinct, so Strict Triangle Inequality works. Thus, $2^{r-1}\cdot ord(\epsilon)=
    (2^r+2^{r-2})ord(x)$ (which is the highest). So
    \begin{displaymath}
        ord(\epsilon)=2^r+2^{r-2}.
    \end{displaymath}
    Now note that $ord(xy)=ord(x)+ord(y)=2^{r-1}+2^{r-1}+2^{r-2}=2^r+2^{r-2}=ord(\epsilon).$
    \begin{rem23}
        $ord(xy)\ne ord(\epsilon)$ if $q\ne 2$, so our
        considerations are essentially valid only for $q=2$.
    \end{rem23}
    Let us do the summing up again. Consider
    $\theta:=\epsilon+xy=x^3+y^2+xy$.
    \begin{eqnarray}
        \theta^{2^{r-1}}=\epsilon^{2^{r-1}}+x^{2^{r-1}}y^{2^{r-1}}=|using\ (2)|= \nonumber\\
        \epsilon^{2^{r-1}}+x^{2^{r-1}}(y^{2^{r-2}}+\dots+y^2+y+x^{2^{r-1}+2^{r-2}}+\dots+x^3)= \nonumber\\
        y+x^{2^r+2^{r-2}}+\dots+x^{2^r+2}+x^{2^{r-1}+1}+x^{2^{r-2}+1}
        \nonumber\\
        +\dots+x^3+x^{2^{r-1}}y^{2^{r-2}}+\dots+x^{2^{r-1}}y^2+x^{2^{r-1}}y+\nonumber\\
        x^{2^r+2^{r-2}}
        +\dots+x^{2^r+2}+x^{2^r+1}+x^{2^{r-1}+2^{r-2}+2^{r-3}}+\dots+x^{2^{r-1}+3} \nonumber\\
        =y+x^{2^{r-1}+1}+
        x^{2^{r-2}+1}+\dots+x^3+x^{2^{r-1}}y^{2^{r-2}}+\nonumber\\
        \dots+x^{2^{r-1}}y^2+x^{2^{r-1}}y
        +x^{2^r+1}+x^{2^{r-1}+2^{r-2}+2^{r-3}}+\dots+x^{2^{r-1}+3}.
        \nonumber
    \end{eqnarray}
    For Strict Triangle Inequality to work we need that
    the highest orders are not duplicated (if some lower orders
    are duplicated , we can sum corresponding functions and a
    resulting function will have either the same order, i.e.
    duplication is removed, or the lower order, so we can repeat
    our procedure, and so on). In order to find the highest order
    among orders of our summands we need to compare the orders of
    $x^{2^r+1}$ and $x^{2^{r-1}}y^{2^{r-2}}:
    ord(x^{2^r+1})=(2^r+1)2^{r-1};
    ord(x^{2^{r-1}}y^{2^{r-2}})=2^{r-1}\cdot
    2^{r-1}+2^{r-2}(2^{r-1}+2^{r-2})=2^{r-2}(2^r+2^{r-2}).$
    Now
    $2^{r-1}(2^r+1)=2^{r-2}(2^{r+1}+2)=2^{r-2}(2^r+2^{r-1}+2^{r-1}+2)>2^{r-2}(2^r+2^{r-2}).$
    So $ord(\theta^{2^{r-1}})=(2^r+1)\cdot 2^{r-1} \Rightarrow
    ord(\theta)=2^r+1$.
    \end{proof}
    The question of calculating of orders of functions was also studied in \cite{11}.
    Obviously, $ord(\theta)=2^r+1$ is not a linear
    combination of $2^{r-1}$ and $2^{r-1}+2^{r-2}$. It turns out that this is all what we
    need in order to construct the Weierstrass semigroup of
    $Q_{\infty}$ (we denote it as $WS(Q_{\infty}))$. Namely, the
    following holds
    \begin{th24}
        $WS(Q_{\infty})=\mathbb{N}\cdot 2^{r-1}+\mathbb{N}\cdot
        (2^{r-1}+2^{r-2})+\mathbb{N}\cdot (2^r+1).$

    \end{th24}\label{th24}
    This theorem can be proven via direct computations (cf. \cite{10}). But we will use so-called telescopic semigroups,
    which will yield a short and elegant proof of the theorem. First, let us define what a telescopic semigroup is and
    give a result that we will use in the proof.
    \begin{def25}(Definition 5.31, \cite{8})
        Let $(a_1,\dots,a_k)$ be a sequence of positive integers with greatest common divisor 1. Define
        \begin{displaymath}
            d_i=gcd(a_1,\dots,a_i) \textrm{ and } A_i=\{a_1/d_i,\dots,a_i/d_i\}
        \end{displaymath}
        for $i=1,\dots,k$. Let $d_0=0$. Let $\Lambda_i$ be the semigroup generated by $A_i$. If $a_i/d_i\in
        \Lambda_{i-1}$ for $i=2,\dots,k$, then the sequence $(a_1,\dots,a_k)$ is called \emph{telescopic}. A
        semigroup is called telescopic if it is generated by a telescopic sequence.
    \end{def25}
    \begin{def26}(Section 5.1, \cite{8})
        Let $\Lambda$ be a semigroup. The number of gaps is denoted by $g=g(\Lambda)$. If $g<\infty$, then there
        exists an $n\in\Lambda$ such that if $x\in\mathbb{N}_0$ and $x\ge n$, then $x\in\Lambda$. The conductor of
        $\Lambda$ is the smallest $n\in\Lambda$ such that $\{x\in\mathbb{N}_0|x\ge n\}$ is contained in $\Lambda$,
        denoted by $c=c(\Lambda)$. So $c-1$ is the largest gap of $\Lambda$ if $g>0$. A semigroup is called
        \emph{symmetric} if $c=2g$.
    \end{def26}
    Now we are ready to give the result.
    \begin{prop27}(Proposition 5.35, \cite{8})
        Let $\Lambda_k$ be the semigroup generated by the telescopic sequence $(a_1,\dots,a_k)$. Then
        \begin{eqnarray}
            c(\Lambda_k)-1=d_{k-1}(c(\Lambda_{k-1})-1)+(d_{k-1}-1)a_k=\sum_{i=1}^k(d_{i-1}/d_i-1)a_i,\nonumber\\
            g(\Lambda_k)=d_{k-1}g(\Lambda_{k-1})+(d_{k-1}-1)(a_k-1)/2=c(\Lambda_k)/2.\nonumber
        \end{eqnarray}
        So telescopic semigroups are symmetric. Here we put $d_0=0$.
    \end{prop27}
    \begin{proof}(of Theorem 2.4)
        Let $\Lambda(r)=<2^{r-1},2^{r-1}+2^{r-2},2^r+1>, r\ge 3$ be a semigroup generated by
        $2^{r-1}=:a_1, \: 2^{r-1}+2^{r-2}=:a_2$, and $2^r+1=:a_3$ for given $r\ge 3$. It is clear that
        $gcd(a_1,a_2,a_3)=1$. Let us check the definition of a telescopic semigroup:
        \begin{eqnarray}
            d_1=gcd(a_1)=2^{r-1}, A_1=\{1\}, \Lambda_1=\mathbb{N}; \nonumber\\
            d_2=gcd(a_1,a_2)=2^{r-2}, A_2=\{2,3\}, \Lambda_2=<2,3>; \nonumber\\
            d_3=gcd(a_1,a_2,a_3)=1, A_3=\{a_1,a_2,a_3\}, \Lambda_3=\Lambda(r); \nonumber
        \end{eqnarray}
        It is clear that $A_2=\{2,3\}\subseteq\mathbb{N}=\Lambda_1$. Also $2^{r-1}\in\Lambda_2=<2,3>$, and
        $2^{r-1}+2^{r-2}\in\Lambda_2$. Finally, $2^r+1=2\cdot(2^{r-1}-1)+3\cdot 1\in <2,3>$. This means that
        $\Lambda(r)$  is a telescopic semigroup. Let us apply Proposition 2.7 to $\Lambda(r)$. We obtain:
        \begin{displaymath}
            c(\Lambda(r))=(d_0/d_1-1)a_1+(d_1/d_2-1)a_2+(d_2/d_3-1)a_3+1.
        \end{displaymath}
        So that
        \begin{displaymath}
            c(\Lambda(r))=-a_1+a_2+(2^{r-2}-1)a_3+1=2^{2r-2}-2^{r-1}.
        \end{displaymath}
        As telescopic semigroups are symmetric, we have:
        \begin{displaymath}
            g(\Lambda(r))=c(\Lambda(r))/2=2^{2r-3}-2^{r-2}.
        \end{displaymath}
        Note, that $g(\Lambda(r))=g(WS(Q_{\infty}))$ per Theorem 1.2. Considering the fact that
        $\Lambda(r)\subseteq WS(Q_{\infty})$  we conclude that $\Lambda(r)=WS(Q_{\infty})$.
    \end{proof}
    As a straightforward, but very important corollary, we have
    \begin{th28}
        $\mathcal{L}(sQ_{\infty})=<x^i y^j \theta^k>_{i,j,k}$,
        where $\theta=x^3+y^2+xy$ and $i\cdot 2^{r-1}+j(2^{r-1}+2^{r-2})+k\cdot(2^r+1)\le
        s$; $\:i,k\ge 0, j\in\{0,1\}$.
    \end{th28}\label{th28}
    \begin{proof}
        This is easily seen as
        $\dim\mathcal{L}(sQ_{\infty})=|WS(Q_{\infty})\cap\{0,1,\dots,s\}|$,
        and functions of the form $x^i y^j \theta^k$ as above are
        linearly independent, because they have different orders
        at $Q_{\infty}$.
    \end{proof}
    In the next section we are going to show how this theorem
    applies to codes.

    \begin{rem29}
        It can be shown (cf. \cite{10}, proof of Theorem 1.3) that the numbers
        $i\cdot 2^{r-1}+j(2^{r-1}+2^{r-2})+k\cdot(2^r+1)$ are all different provided that $i,k\ge 0, j\in\{0,1\}$.
    \end{rem29}\label{rem29}

    \section{Application to GH-codes}
    In coding theory (\cite{1}, \cite{3}) Hermitian codes have
    taken a special place, as this class of codes provides
    interesting and non-trivial examples of Goppa codes. These
    codes are over $\mathbb{F}_{q^2}$, they are not too short
    compared with the size of the alphabet, and their parameters
    $k$ (dimension) and $d$ (minimum distance) are fairly good. In
    addition there is an efficient way to produce generator matrices
    for these codes.
    \begin{def31}(\cite{1}, Definition VII.4.1)
        For $s\in\mathbb{N}$ we define
        \begin{displaymath}
            H_s := C_{\mathcal{L}}(D,sQ_{\infty}),
        \end{displaymath}
        where
        \begin{displaymath}
            D := \sum_{\beta^q+\beta=\alpha^{q+1}} P_{\alpha,\beta}
        \end{displaymath}
        is the sum of all places of degree one except $Q_{\infty}$
        of the Hermitian function field
        $H/\mathbb{F}_{q^2}$ (cf. Proposition 1.1). The codes
        $H_s$ are called \emph{Hermitian} codes.
    \end{def31}
    All the basic facts on performance of Hermitian code can be
    found in \cite{1}.
    \\
    \indent
    We will now treat the Generalized Hermitian codes.
    \begin{def32}
        For $s\in\mathbb{N}$ we define
        \begin{displaymath}
            GH_s := C_{\mathcal{L}}(D,sQ_{\infty}),
        \end{displaymath}
        where
        \begin{displaymath}
            D := \sum_{\beta^{q^{r-1}}+\dots+\beta=
            \alpha^{q^{r-1}+q^{r-2}}+\dots+\alpha^{1+q}} P_{\alpha,\beta}
        \end{displaymath}
        is the sum of all places of degree one except $Q_{\infty}$
        of the Generalized Hermitian function field
        $GH/\mathbb{F}_{q^r}$ (cf. Theorem 1.2). We call the codes
        $GH_s$ \emph{Generalized Hermitian} codes. They are
        Hermitian codes for $r=2$.
    \end{def32}
    GH-codes are codes of length $n=q^{2r-1}$ over
    $\mathbb{F}_{q^r}$. For $t\le s$ we have $GH_t\subseteq GH_s$.
    Now if $s-q^{2r-1}>2g-2 \Rightarrow
    s>q^{2r-1}+q^{2r-2}-q^{r-1}-2$. Riemann-Roch Theorem and
    Theorem II.2.2 from \cite{1} yield $\dim
    GH_s=\dim(sQ_{\infty}) -
    dim(sQ_{\infty}-D)=(s+1-g)-(s+1-g-q^{2r-1})=q^{2r-1}=n$, which
    is trivial. So GH-codes are interesting for $0<s\le
    q^{2r-1}+q^{2r-2}-q^{r-1}-2$.
    \\
    \indent
    Denote $\mathcal{S}(s):=WS(Q_{\infty})\cap\{0,1,\dots,s\}$. We
    have
    $|\mathcal{S}(s)|=s+1-g=s+1-\frac{q^{r-1}(q^{r-1}-1)}{2}$ for
    $s\ge 2g-1=q^{r-1}(q^{r-1}-1)-1$. As before, we will restrict
    ourselves to the case $q=2$. From section 2 we have
    \begin{displaymath}
        \mathcal{S}(s)=\{l\le s|l=i\cdot 2^{r-1}+j\cdot
        (2^{r-1}+2^{r-2})+k\cdot (2^r+1); i,k\ge 0\, j=0,1\}.
    \end{displaymath}
    Some insight on parameters of the code $GH_s$ over
    $\mathbb{F}_{2^r}$ gives the following:
    \begin{prop33}
        Suppose $0<s\le 2^{2r-1}$. Then
        \\
        \indent
        (a) The dimension of $GH_s$ is given by
        \begin{equation}
            \dim GH_s=|\mathcal{S}(s)|.
            \label{1-3-1}
        \end{equation}
        For $2^{2r-2}-2^{r-1}-1<s<2^{2r-1}$, we have
        \begin{equation}
            \dim GH_s = s+1-2^{r-2}(2^{r-1}-2).
            \label{1-3-2}
        \end{equation}
        \indent
        (b) The minimum distance $d$ of $GH_s$ satisfies
        \begin{equation}
            d\ge 2^{2r-1}-s.
            \label{1-3-3}
        \end{equation}
    \end{prop33}\label{prop33}
    \begin{proof}
        \indent
        (a) For $0<s<2^{2r-1}$, Corollary II.2.3 (\cite{1})
        gives $\dim
        GH_s=\dim\mathcal{L}(sQ_{\infty})=|\mathcal{S}(s)|$. The
        formula (\ref{1-3-2}) is straightforward.
        \\
        \indent
        (b) Inequality (\ref{1-3-3}) follows from Theorem II.2.2
        (\cite{1}).
    \end{proof}
    Of course, this proposition remains valid for arbitrary $q$,
    but for $s\le 2^{2r-2}-2^{r-1}-1$ the description of
    $|\mathcal{S}(s)|$, which we obtained in section 2, is crucial.
    \\
    \indent
    Let us now present a generator matrix for the GH-codes over
    $\mathbb{F}_{2^r}$. We fix an ordering of the set
    $T:=\{(\alpha,\beta)\in
    \mathbb{F}_{2^r}\times\mathbb{F}_{2^r}|\beta^{2^{r-1}}+\dots+\beta=\alpha^{2^{r-1}+2^{r-2}}
    +\dots+\alpha^3\}.$ For $l=i\cdot 2^{r-1}+j(2^{r-1}+2^{r-2})+k\cdot(2^r+1)s;\: i,k\ge 0, j=0,1$
    we define a vector
    \begin{displaymath}
        u_l := (\alpha^i \beta^j
        (\alpha^3+\beta^2+\alpha\beta)^k)_{(\alpha,\beta)\in T}\in
        (\mathbb{F}_{2^r})^{2^{2r-1}}.
    \end{displaymath}
    As a corollary of Theorem 2.8 and Corollary II.2.3
    (\cite{1}) we have:
    \begin{prop34}
        Suppose that $0<s<2^{2r-1}$ and let $k:=|\mathcal{S}(s)|$.
        Then the $k\times 2^{2r-1}$ matrix
        \begin{equation}
            GHM_s := (u_l)_{l\in\mathcal{S}(s)}
            \label{1-3-4}
        \end{equation}
        is a generator matrix for $GH_s$.
    \end{prop34}
    Now we will show how an estimate from Proposition 3.3 can be improved by applying results from \cite{9}.
    For this we define
    \begin{equation}
        C'_s=\big(C_{\mathcal{L}}(D, \rho_sQ_{\infty})\big)^{\perp}=C_{\Omega}(D,\rho_sQ_{\infty}),
        \label{10}
    \end{equation}
    where $WS(Q_{\infty})=(\rho_i)_{i\in\mathbb{N}}$ is a non-gap sequence of $Q_{\infty}$.
    \\
    \indent
    For these codes a designed Feng-Rao distance $\delta_{FR}(s)$ can be defined (for definition cf. \cite{9}).
    Without going deeply into details we only state that:
    \begin{displaymath}
        d(C'_s)\ge \delta_{FR}(s)
    \end{displaymath}
    (Theorem 2.5, \cite{9}), and
    \begin{displaymath}
        \delta_{FR}(s)\ge\delta_{\Gamma}(s)
    \end{displaymath}
    where $\delta_{\Gamma}(s)$ is a Goppa designed distance of $C'_s$ (Corollary 3.9, \cite{9}).
    Note that the estimate in Proposition 3.3 is given via this designed distance.
    \\
    \indent
    In \cite{9} C.Kirfel and R.Pellikaan give some estimates on $\delta_{FR}$ for the case when a Weierstrass semigroup
    is telescopic. As  this is the case in our situation we can apply these results. First we quote:
    \begin{th35}(Theorem 6.10, \cite{9})
        Let the semigroup of non-gaps at $P$ ($P=Q_{\infty}$ in (\ref{10})) be generated by the telescopic sequence
        $(a_1,\dots,a_k)$. Suppose $a_k=max(A_k)$ and $d_{k-1}=gcd(a_1,\dots,a_{k-1})>1$. Let $(\rho_i)$ be the non-gap
        sequence at $P$. For codes $C(r)=C_{\Omega}(D,\rho_rP)$ we have
        \begin{displaymath}
            \delta_{FR}(r)=min\{\rho_t|\rho_t\ge r+1-g\},
        \end{displaymath}
        if $3g-2-(d_{k-1}-1)a_k<r\le 3g-2$ and $g\le r$.
    \end{th35}
    \begin{th36}(Theorem 6.11, \cite{9})
        Let the semigroup of non-gaps at $P$ be generated by the telescopic sequence
        $(a_1,\dots,a_k)$. Suppose $a_k=max(A_k)$. If
        \begin{displaymath}
            (j-1)a_k<\rho_{r+1}\le ja_k\le (d_{k-1}-1)a_k
        \end{displaymath}
        then
        \begin{displaymath}
            \delta_{FR}(r)=j+1.
        \end{displaymath}
    \end{th36}
    A direct application to our situation yields:
    \begin{prop37}
        Let $C'_s$ be defined as above. The the following holds:
        \begin{displaymath}
            \delta_{FR}(s)=min\{\rho_t|\rho_t\ge s+1-g\},
        \end{displaymath}
        if $3g-2-(2^{r-2}-1)(2^r+1)<s\le 3g-2$ and $g\le s$, where $g=2^{r-2}(2^{r-1}-1), r\ge 3$.
    \end{prop37}
    \begin{proof}
        We have (cf. the proof of Theorem 2.4 ): $k=3, d_{k-1}=d_2=2^{r-2}>1, a_3=max(A_3)=2^r+1$.
    \end{proof}
    \begin{prop38}
        In the notation as above, if
        \begin{displaymath}
            (j-1)(2^r+1)<\rho_{s+1}\le j(2^r+1)\le (2^{r-2}-1)(2^r+1)
        \end{displaymath}
        then
        \begin{displaymath}
            \delta_{FR}(s)=j+1.
        \end{displaymath}
    \end{prop38}
    \begin{ex39}
        Let us consider the case $q=2,r=3$, then $g=6$. From Proposition 3.7  we have
        $\delta_{FR}(s)=min\{\rho_t|\rho_t\ge s-5\}$, if $7<s\le 16$. The following table lists $s, \delta_{FR}(s)$ ,
        and $\delta_{\Gamma}(s)$ for $s=8,\dots,16$. Where $\delta_{FR}(s)>\delta_{\Gamma}(s)$ a bold font
        is used.
        \\
            \begin{tabular}{|c|c|c|}
                \hline
                \textbf{$s$} & \textbf{$\delta_{FR}(s)$} &
                \textbf{$\delta_{\Gamma}(s)$} \\
                \hline
                \bf{8} & \bf{4} & \bf{3} \\
                \hline
                9 & 4 & 4 \\
                \hline
                \bf{10} & \bf{6} & \bf{5} \\
                \hline
                11 & 6 & 6 \\
                \hline
                \bf{12} & \bf{9} & \bf{7} \\
                \hline
                \bf{13} & \bf{9} & \bf{8} \\
            \hline
                14 & 9 & 9 \\
            \hline
                15 & 10 & 10 \\
            \hline
                \bf{16} & \bf{12} & \bf{11} \\
            \hline
            \end{tabular}
            \\
            \indent
        The case $s=16$ is of particular interest, see Section 5.
    \end{ex39}

    \section{Duality property}
    In this section we want to establish a duality property for GHC, which turns out to generalize the one of HC. First
    of all, let us recall the corresponding result for Hermitian codes.
    \begin{prop41}
        The dual code of $H_s$ is
        \begin{displaymath}
            H_s^{\perp}=H_{q^3+q^2-q-2-s}.
        \end{displaymath}
        Hence $H_s$ is self-orthogonal if $2s\le q^3+q^2-q-2$, and $H_s$ is self-dual if and only if
        $s=(q^3+q^2-q-2)/2$.
    \end{prop41}
    Now we will formulate an analogous result in a more general setting and then apply it to GHC.
    \\
    \indent
    Consider a curve $\mathcal{G}$ over $\mathbb{F}_{q^r}$ given by an equation
    \begin{displaymath}
        (f(y))^q+y=g(x),
    \end{displaymath}
    $f(T), g(T)\in\mathbb{F}_q[T]$. Suppose that $\mathcal{G}$ is absolutely irreducible. Denote a function field of
    $\mathcal{G}$ by $\mathcal{F}$. Let $N=N(\mathcal{F})$ and $g=g(\mathcal{F})$ denote the number of rational
    points and the genus of $\mathcal{F}$ resp. Suppose further that the pole $P_{\infty}$ of $x$ in
    $\mathbb{F}_{q^r}(x)$ has a unique extension $Q_{\infty}\in\mathbb{P}_{\mathcal{F}}$, and $Q_{\infty}|P_{\infty}$ is
    totally ramified. From this it follows that $Q_{\infty}$ is a place of $\mathcal{F}/\mathbb{F}_{q^r}$ of degree one
    and $(x)_{\infty}=q\cdot deg f(T)$ (cf. the proof of Proposition 1.4). Finally, assume that for each
    $\alpha\in\mathbb{F}_{q^r}$, there are $q\cdot deg f(T)$ elements $\beta\in\mathbb{F}_{q^r}$ such that
    $(f(\beta))^q+\beta=g(\alpha)$ , and for all such pairs $(\alpha,\beta)$ there is a unique place $P_{\alpha,\beta}$
    of degree one with $x(P_{\alpha,\beta})=\alpha,y(P_{\alpha,\beta})=\beta$.
    \\
    \indent
    Consider a family of codes
    \begin{displaymath}
        \mathcal{C}_l=C_{\mathcal{L}}(D,lQ_{\infty}),
    \end{displaymath}
    where $D=\sum_{(f(\beta))^q+\beta=g(\alpha)}P_{\alpha,\beta}$. We will be interested in the case, when
    $0\le l\le N+2g-3$. Note that HC and GHC are $\mathcal{C}_l$-codes for a special choice of the  curve $\mathcal{G}$.
    \begin{th42}
        The dual code of $\mathcal{C}_l$ is
        \begin{displaymath}
            \mathcal{C}_l^{\perp}=\mathcal{C}_{N+2g-3-l}.
        \end{displaymath}
    \end{th42}
    \begin{proof}
        First of all we will need the following lemma.
        \begin{lem43}
            The divisor of the differential $dx$ is
            \begin{displaymath}
                (dx)=(2g-2)Q_{\infty}.
            \end{displaymath}
            (for an analogous result for Hermitian codes cf. Lemma VI.4.4(d),\cite{1})
        \end{lem43}
        \begin{proof}
            From Remark IV.3.7(c), \cite{1} we have
            \begin{equation}
                (dx)=-2(x)_{\infty}+Diff(\mathcal{F}/\mathbb{F}_{q^r}(x)),
                \label{11}
            \end{equation}
            where $Diff(\mathcal{F}/\mathbb{F}_{q^r}(x))$ is a \emph{different} of $\mathcal{F}/\mathbb{F}_{q^r}(x)$.
            We know that $(x)_{\infty}=q\:deg f(T)\cdot Q_{\infty}$, so we have to calculate the different. We
            need another lemma (Theorem III.5.10(a), \cite{1}):
            \begin{lem44}
                Suppose $F'=F(y)$ is a finite separable extension of a function field $F$ of degree
                $[F':F]=n$. Let $P\in\mathbb{P}_F$ be such that the minimal polynomial $\phi(T)$ of $y$ over
                $F$ has coefficients in the valuation ring $\mathcal{O}_P$ of $P$(i.e. $y$ is integral over
                $\mathcal{O}_P$), and let $P_1,\dots,P_m\in\mathbb{P}_{F'}$ be all places of $F'$ lying over
                $P$. Then
                \begin{displaymath}
                    d(P_i|P)\le v_{P_i}(\phi '(y)) \textrm{ for } 1\le i\le m,
                \end{displaymath}
                where $d(P_i|P)$ is a \emph{different exponent} of $P_i$ over $P$.
            \end{lem44}
            Now, $\forall\mathbb{P}_{\mathbb{F}_{q^r}(x)}\ni P\ne P_{\infty}: g(x)\in\mathcal{O}_P$, so
            $\phi (T)=(f(T))^q+T-g(x)\in\mathcal{O}_P[T]$. Next, $\phi '(y)=1$. So
            $\forall\mathbb{P}_{\mathcal{F}}\ni P'|P: d(P'|P)\le v_{P'}(1)=0$. By definition of $d(P'|P)$ we have
            that $d(P'|P)\ge 0$. It follows that $d(P'|P)=0$. As $P_{\infty}$ is totally ramified,
            we obtain that $Diff(\mathcal{F}/\mathbb{F}_{q^r}(x))=a\cdot Q_{\infty}$. By Hurwitz genus formula
            (Theorem III.4.12, \cite{1})
            \begin{displaymath}
                a=deg Diff(\mathcal{F}/\mathbb{F}_{q^r}(x))=2g-2+2q\cdot deg f(T).
            \end{displaymath}
            Collecting all the above, (\ref{11}) yields:
            \begin{displaymath}
                (dx)=(-2q\cdot deg f(T)+2q\cdot deg f(T)+2g-2)Q_{\infty}=(2g-2)Q_{\infty}.
            \end{displaymath}
        \end{proof}
        Now we can rewrite the proof of Proposition VII.4.2, \cite{1} for our situation. Consider the element
        \begin{displaymath}
            z:=\prod_{\alpha\in\mathbb{F}_{q^r}}(x-\alpha)=x^{q^r}-x.
        \end{displaymath}
        $z$ is a prime element for all places $P_{\alpha,\beta}\le D$, and its principal divisor is
        $(z)=D-(N-1)Q_{\infty}$. Since $dz=d(x^{q^r}-x)=-dx$, the differential $dz$ has the
        divisor $(dz)=(dx)=(2g-2)Q_{\infty}$ due to Lemma 4.3. Now Theorem II.2.8 and Proposition VII.1.2, \cite{1} imply
        \begin{eqnarray}
            \mathcal{C}_l^{\perp}=C_{\Omega}(D,lQ_{\infty})=
            C_{\mathcal{L}}(D,D-lQ_{\infty}+(dz)-(z))=\nonumber\\
            C_{\mathcal{L}}(D,((N-1)+2g-2-lQ_{\infty})=\mathcal{C}_{N+2g-3-l}.\nonumber
        \end{eqnarray}
    \end{proof}
    It is clear the Proposition 4.1 is a corollary of Theorem 4.2. A result for GHC looks as follows:
    \begin{cor45}
        The dual code of $GH_s$ is
        \begin{displaymath}
            GH_s^{\perp}=GH_{q^{2r-1}+2g-2-s}=GH_{q^{2r-1}+q^{r-1}(q^{r-1}-1)-2-s}.
        \end{displaymath}
        Hence $GH_s$ is self-orthogonal if $2s\le q^{2r-1}+q^{r-1}(q^{r-1}-1)-2$, and $GH_s$ is self-dual (this case
        can only occur if $q$ is a power of 2) if and only if $s=(q^{2r-1}+q^{r-1}(q^{r-1}-1)-2)/2$.
    \end{cor45}

    \section{Computational results}
    Here we demonstrate some computational results on GH-codes over
    $\mathbb{F}_{2^3}$. The codes (their generator matrices) were computed
    using SINGULAR computer algebra system \cite{4},\cite{7} the minimum
    distance was computed in GAP computer algebra system
    \cite{5}.

    In the table below $d_{rec}$ is a record value for $d$ for
    given $n=32$ and $k$. These are taken from Brouwer's table (\cite{6}) for
    the linear codes over $\mathbb{F}_{8}$.
    \\
    \begin{tabular}{|c|c|c|}
        \hline
        \textbf{$k$} & \textbf{$d_{rec}$} &
        \textbf{$d$} \\
        \hline
        6 & 22 & 22 \\
        \hline
        7 & 20 & 20 \\
        \hline
        8& 20 & 19 \\
        \hline
        9& 18 & 18 \\
        \hline
        10& 17 & 17 \\
        \hline
        11& 16 & 16 \\
        \hline
    \end{tabular}
    \\
    \indent
    When discussing estimates on Feng-Rao designed distance in section 3, we saw for k=16, $\delta_{FR}(16)\ge 12$.
    Thus we obtained $[32,16,\ge 12]$-code over $\mathbb{F}_{8}$ (in a view of the duality property).
    This yields a new record, which is cited in Brouwer's table.

    \section{Conclusion}
    In this paper we studied generalization of Hermitian function field proposed by A.Garcia and H.Stichtenoth.
    We calculated a Weierstrass semigroup of the point at infinity for the case $q=2, r\ge 3$. It turned out that
    unlike Hermitian case, we have already three generators for the semigroup. We then applied this result to
    codes, constructed on generalized Hermitian function fields. Further, we applied results of C.Kirfel and
    R.Pellikaan to estimating a Feng-Rao designed distance for GH-codes, which improved on Goppa designed
    distance. Next, we studied the question of codes dual to GH-codes. We identified that the duals are also GH-codes
    and gave an explicit formula. We concluded with some computational results. In particular, a new record-giving
    $[32,16,\ge 12]$-code over $\mathbb{F}_{8}$ was presented. As a further work we see studying a structure of the
    Weierstrass semigroup for other values of $q$. It could also be interesting to apply a theory of generalized weights
    to GH-codes.

    \section{Acknowledgement}
    The author would like to express his appreciation to R.Pellikaan (Technical University of Eindhoven, Netherlands)
    and to S.A.Ovsienko (Kyiv National University, Ukraine) for their useful suggestions, remarks, and support.

\end{document}